\documentclass[final,3p,times,twocolumn]{elsarticle}
\biboptions{comma,sort&compress}
\usepackage{here}
\usepackage{graphicx}
\usepackage{epsfig}
\usepackage{color}
\usepackage{slashed}%\slash{符号}

\definecolor{mycrimson}{rgb}{0.50,0.00,0.00}
\definecolor{myblue}{rgb}{0.00,0.00,0.63}

\def\nin{\noindent}
\def\bea{\begin{eqnarray}}
\def\eea{\end{eqnarray}}

\newcommand{\intd}{{1\over i}\int{\texttt{d}^dk\over(2\pi)^d}}
\journal{Nuc. Phys. (Proc. Suppl.)}

\begin{document}

\begin{frontmatter}



\title{Pion-nucleon elastic scattering amplitude within covariant baryon chiral perturbation theory up to $O(p^4)$ level}

\author[label1]{Yun-hua Chen}
%\ead{huacyh@pku.edu.cn}
\author[label1]{De-liang Yao\corref{cor1}}
\ead{yaodeliang@pku.edu.cn}
\author[label1]{H.~Q. Zheng}
%\ead{zhenghq@pku.edu.cn}
\address[label1]{ Department of Physics and State Key Laboratory of
Nuclear Physics and Technology, Peking University, Beijing 100871,
China P.R.}
\cortext[cor1]{Speaker}
 %\author{Yun-hua Chen, De-liang Yao, H.~Q. Zheng}

% \author[label1,label2]{Marina Nielsen,\corref{label3}}

%\ead{mnielsen@if.usp.br}
%\author{}

%\address{}

\begin{abstract}
\noindent The $O(p^4)$ calculation on pion-nucleon elastic
scattering amplitude in EOMS scheme within covariant baryon chiral
perturbation theory is reviewed. Numerical fits to partial wave
amplitudes up to $\sqrt{s}=1.13$GeV and $1.20$GeV are performed and the results are
compared with previous studies.

\end{abstract}


\end{frontmatter}

% \linenumbers

\section{Introduction}
%\label{}
\noindent
Many efforts have been made in studying $\pi$-$N$ scatterings at
low energies. However, unlike the successfulness of chiral
perturbation theory in pure mesonic sector, a chiral expansion in
$\pi$-$N$ scattering amplitude suffers from the power counting
breaking (PCB) problem in the traditional subtraction
$\overline{MS}-1$ scheme.~\cite{gasser} Many proposals have been
made to treat this problem, e.g., heavy baryon chiral perturbation
theory~\cite{HB}, infrared regularization scheme~\cite{becherleutwyler},
extended on mass shell (EOMS) scheme~\cite{EOMS}, etc.. The EOMS
scheme provides a good solution to the PCB problem, e.g., see~\cite{Geng}, in the sense that
it faithfully respects the analytic structure of the original
amplitudes and being scale independent.

In this talk we will present our work on the $O(p^3)$ and $O(p^4)$
calculation on $\pi$-$N$ scattering amplitude in EOMS scheme and will
compare it with previous results in the literature.

\nin
%%%%%%%%%%%%

%%%%%%%%%%%%
\section{NNLO and NNNLO calculations}
{\scriptsize
\begin{table*}[ht]
\caption{Fitting results at ${\cal O}(p^3)$ and ${\cal O}(p^4)$. Given for comparison are results from~\cite{oller1,oller3}. The $c_i$, $d_j$ and $e_k$ have, respectively, units of $\rm{GeV}^{-1}$, $\rm{GeV}^{-2}$ and $\rm{GeV}^{-3}$. In ${\cal O}(p^4)$ fits, the fitted $c_i$ here should be understood as $\hat{c_i}$: $\hat{c_1}=c_1-2M^2\left(e_{22}-4e_{38}\right)$, $\hat{c_2}=c_2+8M^2\left(e_{20}+e_{35}\right)$, $\hat{c_3}=c_3+4M^2\left(2e_{19}-e_{22}-e_{36}\right)$, $\hat{c_4}=c_4+4M^2\left(2e_{21}-e_{37}\right)$.}
\begin{center}
%\newcolumntype{C}{>{\centering\arraybackslash}X}%    定义居中对齐命令
%\newcolumntype{R}{>{\raggedleft\arraybackslash}X}%　定义右对齐命令
%\renewcommand{\arraystretch}{1.}%表格行高
%\hspace{-50pt}
\begin{tabular}{c | r r r r|r r}
\hline
 LEC    &        Fit I-${\cal O}(p^3)$      & Ref.~\cite{oller1}-${\cal O}(p^3)$ & Fit II-${\cal O}(p^3)$ & Ref.~\cite{oller3}-${\cal O}(p^3)$&Fit I-${\cal O}(p^4)$&Fit II-${\cal O}(p^4)$\\
\hline\hline
$c_1$   &       $-1.39\pm0.06$     & $-1.50\pm0.06$   & $-0.81\pm0.03$  & $-1.00\pm0.04$& $-1.09\pm0.06$ & $-0.98\pm0.03$\\
$c_2$   &       $4.00\pm0.09$      & $3.74\pm0.09$    & $1.46\pm0.09$   & $1.01\pm0.04$ & $2.79\pm0.10$  & $1.41\pm0.04$ \\
$c_3$   &       $-6.59\pm0.08$     & $-6.63\pm0.08$   & $-3.09\pm0.12$  & $-3.04\pm0.02$& $-5.32\pm0.14$ & $-3.76\pm0.04$\\
$c_4$   &       $3.91\pm0.04$      & $3.68\pm0.05$    & $2.35\pm0.06$   & $2.02\pm0.01$ & $2.38\pm0.19$  & $1.16\pm0.03$\\
\hline
$d_1+d_2$&      $4.32\pm0.53$      & $3.67\pm0.54$    & $0.78\pm0.09$   &&$6.21\pm0.12$&$2.14\pm0.04$\\
$d_3$   &       $-3.00\pm0.50$     & $-2.63\pm0.51$   & $-0.46\pm0.05$  &&$-6.86\pm0.16$&$-3.88\pm0.05$\\
$d_5$   &       $-0.56\pm0.13$     & $-0.07\pm0.13$   & $-0.16\pm0.04$  &&$0.54\pm0.11$&$1.17\pm0.04$\\
$d_{14}-d_{15}$&$-7.05\pm1.05$     & $-6.80\pm1.07$   & $-0.89\pm0.15$  &&$-11.90\pm0.24$&$-3.96\pm0.08$\\
$d_{18}$&       $-0.74\pm1.41$     & $-0.50\pm1.43$   & $-0.92\pm0.25$  &&$-0.74$(input)&$-0.74$(input)\\
\hline
$e_{14}$   &-&-&-&-&      $3.68\pm0.36$    & $2.62\pm0.09$  \\
$e_{15}$   &-&-&-&-&      $-14.67\pm0.55$  & $-5.15\pm0.13$  \\
$e_{16}$   &-&-&-&-&      $7.15\pm0.35$    & $1.55\pm0.07$  \\
$e_{17}$   &-&-&-&-&      $0.57\pm1.34$    & $13.57\pm0.15$  \\
$e_{18}$   &-&-&-&-&      $3.64\pm1.18$    & $-9.05\pm0.12$  \\
\hline
$h_A$   &             -            &        -         & $2.82\pm0.04$   & $2.87\pm0.04$  &-     & $2.82$(input)\\
\hline
$\chi^2_{d.o.f}$& 0.18             & $0.22$           & 0.35            &0.23            & 0.04 & 0.21\\
\hline
\end{tabular}
\label{table1}
\end{center}
\end{table*}}
%\label{}
\nin
%%%%%%%%%%%%
We start from the following effective lagrangian at $O(p^3)$ level
(extendable to $O(p^4)$~\cite{eff}):
{\small
\bea
\hspace{-20pt}
{\cal L}_{eff}&=&\bar{N}\left\{i\,\slashed{D}-m+\frac{g_A}{2}\,\slashed{u}
\gamma_5+c_i\,{\cal O}^{(2)}_i+d_j\,{\cal O}^{(3)}_j\right\}N+\nonumber\\
&&\frac{f_\pi^2}{4}\langle u^\mu
u_\mu+\chi_+\rangle+\frac{\ell_4}{8}\langle u^\mu
u_\mu\rangle\langle\chi_+\rangle+\frac{\ell_3+\ell_4}{16}\langle\chi_+\rangle^2\
,\nonumber
\eea}
where ${\cal O}^{(2)}$ and ${\cal O}^{(3)}$ are relevant operators
of $O(p^2)$ and $O(p^3)$ respectively, $i\in(1,2,3,4)$ and
 $j\in(1,2,3,5,14,15,16,18)$~\cite{eff}.

Decomposition of $\pi$-$N$ amplitude is standard,%(suited to make chiral expansion)
{\small
\bea
\hspace{-20pt}
T_{\pi N}^{a^\prime a}&=&\delta_{a^\prime
a}T^++\frac{1}{2}[\tau_{a^\prime},\tau_a]T^-,\nonumber\\
T^{\pm}&=&\bar{u}(p^\prime,s^\prime)\left[D^\pm+{i\over
2m_N}\sigma^{\mu\nu}q_\mu^\prime q_\nu B^\pm\right]u(p,s)\ .
\eea}

To carry out the calculation in EOMS scheme one firstly perform
$\overline{MS}-1$ substraction to remove ultraviolet
divergencies, then additional substraction (A.S.) to absorb PCB terms. Taking the nucleon mass renormalization for example,
one has,
{\small
\bea
\label{mNp3}
\hspace{-20pt}
m_N&=&m-4c_1M^2-\frac{3mg^2}{2f^2}\left[\Delta_N-M^2I(m^2)\right]\nonumber\\
&=&\mathring{m}-4c_1^r M^2+\frac{3mM^2g^2}{2f^2}\bar{I}(m^2)\qquad\qquad\quad(\rm{\overline{{MS}}-1})\nonumber\\
&=&\mathring{m}-4\tilde{c_1} M^2+\frac{3m M^2
g^2}{2f^2}\overline{I}(m^2)-{\frac{3m M^2
g^2}{32\pi^2f^2}}\,(\rm{A.S.}),
\eea}
where $\mathring{m}$ is the nucleon mass in chiral limit. The last
term on the $r.h.s.$ of the third equality  is opposite to the PCB term which is absorbed by redefining
$c_1^r$ as: $
\tilde{c_1}=c_1^r-\frac{3g^2m}{128f^2\pi^2}$. Definitions of all
functions appeared here follow from \ref{DefInt}.

Another example is the calculation of the axial-vector coupling
$g_A$:
{\small
\bea
\label{gAp3}
\hspace{-20pt}
g_A&=&g+4d_{16}M^2-\frac{g^3m^2}{32f^2\pi^2}+\frac{g(4-g^2)}{2f^2}\Delta_N-\frac{g(2+g^2)}{2f^2}\Delta_\pi\nonumber\\
\hspace{-20pt}&+&\frac{g^3(2m^2+M^2)}{4f^2}J_N(0)-\frac{g(8-g^2)M^2}{4f^2}I(m^2)-\frac{g^2M^4}{4f^2}I_A(0)\nonumber\\
\hspace{-20pt}&+&\frac{3g^3m^2M^2}{f^2}
\frac{\partial I(s)}{\partial s}I(s)\bigg{|}_{\slashed{p}=m_N}\nonumber\\
\hspace{-20pt}&=&\mathring{g}_A+4d_{16}M^2-\frac{g(2+g^2)}{2f^2}\Delta_\pi+\frac{3g^3m^2M^2}{f^2}\frac{\partial I(s)}{\partial
s}\bigg{|}_{\slashed{p}=m_N}\nonumber\\
\hspace{-20pt}&+&\frac{g^3M^2}{4f^2}J_N(0)-\frac{{g(8-g^2)M^2}}{4f^2}I(m^2)-\frac{g^2M^4}{4f^2}I_A(0)\ ,
\eea }
where $\mathring{g}_A$ is
the axial charge in the chiral limit. Ultraviolet
divergencies are treated by $\overline{MS}-1$ substraction. If we start with
$\mathring{g}_A$, there are no PCB terms to be extracted. The PCB
effects are included in $\mathring{g}_A$. If we start with a bare
$g$, we need to redefine it as, $
\tilde{g}=g^r-\frac{g^3m^2}{16f^2\pi^2},\,\,\,
g^r=g+\frac{g(2-g^2)m^2}{16f^2\pi^2}R$. We prefer the latter hereafter, i.e. starting with bare parameters.

Similar to $m_N$ and $g_A$ renormalization , the calculation of scattering amplitude
up to ${\cal O}(p^3)$ in EOMS scheme is straightforward, if the PCB terms in
functions $D$ and $B$ for loop amplitudes are known,
{\small
\bea
\hspace{-20pt}
D^+_{PCB}&=&\frac {1}{64f^4m\pi^2\sigma^2}\left\{6 g^2 m^2
M^2\sigma^2 +2\sigma^4\right.\nonumber\\
&&\hspace{40pt}+g^4\left[2 m^4\left(10M^4-7M^2t+t^2\right)\right.\nonumber\\
&&\hspace{40pt}\left.\left.+3 m^2\left(3 t-7 M^2\right)\sigma^2+\sigma^4\right]\right\}\ ,\nonumber\\
D^-_{PCB}&=&\frac{g^4m}{64 f^4\pi^2\sigma^2}\left\{\sigma^2\left(t-2M^2
+2\sigma\right)\right.\nonumber\\
&&\hspace{40pt}\left.-2m^2\left(2M^2-t\right)\left(2M^2-t+2\sigma\right)\right\}\ ,\nonumber\\
B^+_{PCB}&=&\frac {g^4m^4}{8f^4\pi^2\sigma^2}\left(2M^2-t
+2\sigma\right)\ ,\nonumber\\
B^-_{PCB}&=&\frac{g^2m^2}{32
f^4\pi^2\sigma^2}\left\{5\sigma^2+g^2\left[4m^2(t-5M^2)+3\sigma^2\right]\right\}\ ,\ \
\eea}
where $\sigma=s-m^2$. After mass and $g_A$ renormalization, the PCB terms above can
be absorbed by redefining $c_i^r$s:
\bea
&&c_1^{r}\rightarrow\widetilde{c_1}=c_1^{r}-\frac{3g^2m}{128F^2\pi^2}\nonumber\\
&&c_2^{r}\rightarrow\widetilde{c_2}=c_2^{r}+\frac{\left(2+g^4\right) m}{32 f^2 \pi ^2},\nonumber\\
&&c_3^{r}\rightarrow\widetilde{c_3}=c_3^{r}-\frac{9 g^4 m}{64 f^2\pi ^2},\nonumber\\
&&c_4^{r}\rightarrow\widetilde{c_4}=c_4^{r}+\frac{g^2\left(5+g^2\right) m}{64 f^2 \pi ^2},
\eea
and the
$\widetilde{c_i}$s are determined by fitting data. Theoretically,
the NNLO amplitudes keep good analytic, correct power counting and
scale-independent properties.

In the following we further extend the above calculation to $O(p^4)$ level:
{\small
\bea\label{p41}
\hspace{-20pt}
m_N&=&m+\cdots-2(8e_{38}+e_{115}+e_{116})M^4\nonumber\\
&&+\frac{3M^2\Delta_\pi}{f^2}\left[\left(2c_1-c_3\right)-\frac{c_2}{d}\right]\ ,
\eea}
{\small
\bea\label{p42}
\hspace{-20pt}
g_A&=&g+\cdots-\frac{2g}{m
f^2}\left\{c_2\left(\frac{4M^2\Delta_\pi+m^2\Delta_N}{d}-M^2I^{(2)}(m^2)\right)\right.\nonumber\\
&&\left.-4m^2\left[(c_3+c_4)I^{(2)}(m^2)+c_4\left(\Delta_\pi-M^2I(m^2)\right)\right]\right\}\ .
\eea}

Only $O(p^4)$ parts are shown explicitly on the $r.h.s.$ of
Eqs.~(\ref{p41}), (\ref{p42}), and ellipses represent lower order contributions given by Eqs.~(\ref{mNp3}), (\ref{gAp3}). It is worth noticing that when obtaining the $O(p^4)$ results, replacement of $m$ in nucleon propagator with $m_2=m-4c_1M^2$, namely making Dyson resummation to renormalize $m$ to $m_2$ first, will simplify calculations greatly~\cite{becherleutwyler}. The $O(p^4)$ part in Eq.~(\ref{p41}) doesn't contribute PCB terms, while the one in Eq.~(\ref{p42}) does and $g^r$ is now redefined as $\tilde{g}=g^r-\frac{g^3m^2}{16f^2\pi^2}+\frac{g m^3}{576f^2\pi^2}\left(9c_2+32c_3+32c_4\right)$.

PCB terms of the
fourth-order loop amplitude read,
{\small
\bea
\hspace{-20pt}
B^+_{PCB}&=&\frac {-m} {576 f^4\pi^2\sigma^3}\left\{ [24{c_4}+(67{c_2}-56{c_3}+96{c_4})g^2]\sigma^4\right.\nonumber\\
&&+32(2{c_2}+17{c_3}
-19{c_4})g^2m^2 M^2\sigma^2\nonumber\\
&&+2(9{c_2}+32{c_3}+32{c_4})g^2m^4\nonumber\\
&&\times\left.\left[4 M^4 \sigma+t^2-t+2 \sigma ^2+M^2
(-4 t+2 \sigma )\right]\right\},\nonumber\\
B^-_{PCB}&=&\frac{m^3}{576 f^4 \pi ^2 \sigma ^3} \left\{\left(9 {c_2}+32{c_3}+16{c_4}\right)\sigma ^3\right.\nonumber\\
&&-2 (9{c_2}+16{c_3}
-28{c_4}) g^2\sigma ^3\nonumber\\
&&\left.+2g^2 m^2(9c_2+32c_3+32c_4)\right.\nonumber\\
&&\left.\times(2 M^2-t)(2
M^2-t+\sigma)\right\},
\eea} and $D^\pm_{PCB}$ terms as well as the
full amplitude are also obtained but are very lengthy, so we will
present it elsewhere.~\cite{CYZ}
%%%%%%%%%%%%%%%%%%%%%%%%%%%%%%%%%%%%%%%%%%%%%%%%%%%%%%
\section{Numerical studies and conclusions}
\begin{figure*}[ht]
\includegraphics[width=0.5\textwidth]{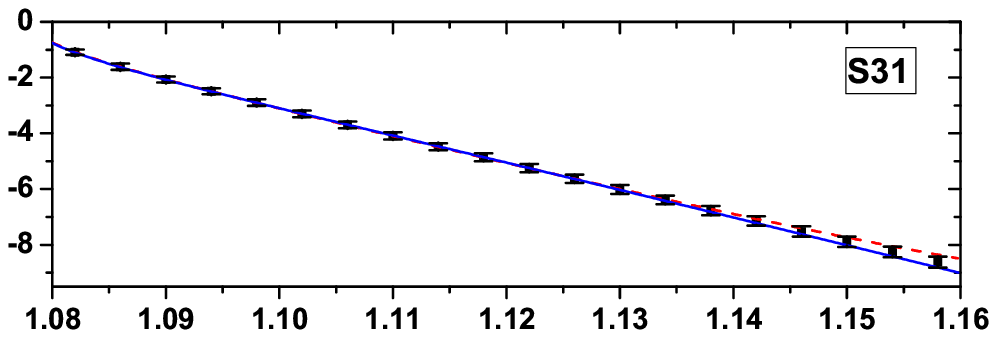}
\vspace{-1.0cm}
\includegraphics[width=0.5\textwidth]{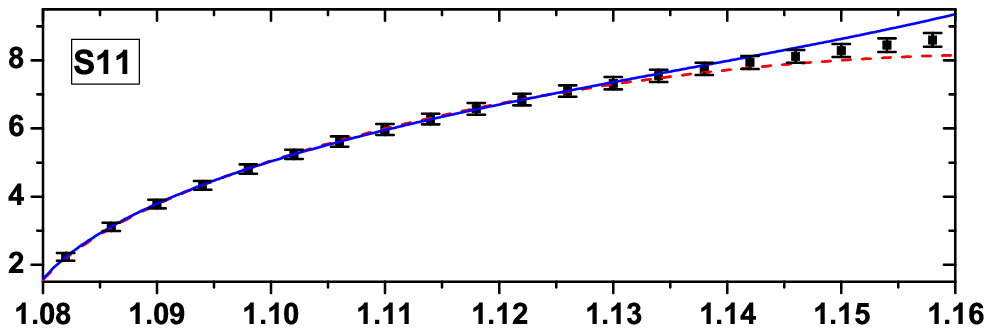}
\vspace{-1.0cm}
\includegraphics[width=0.5\textwidth]{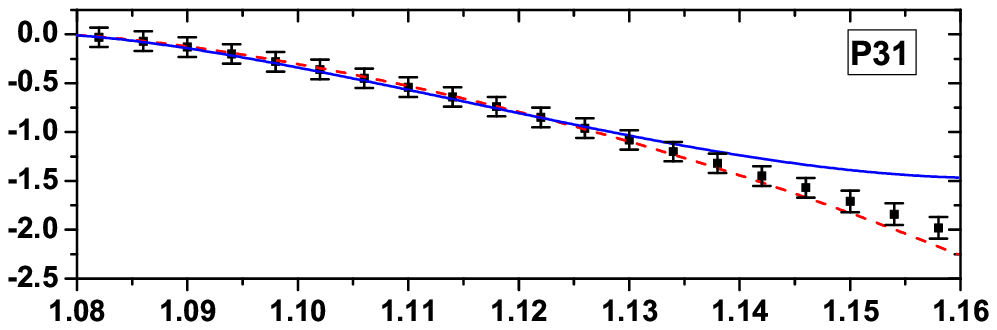}
%\vspace{-1.5cm}
\includegraphics[width=0.5\textwidth]{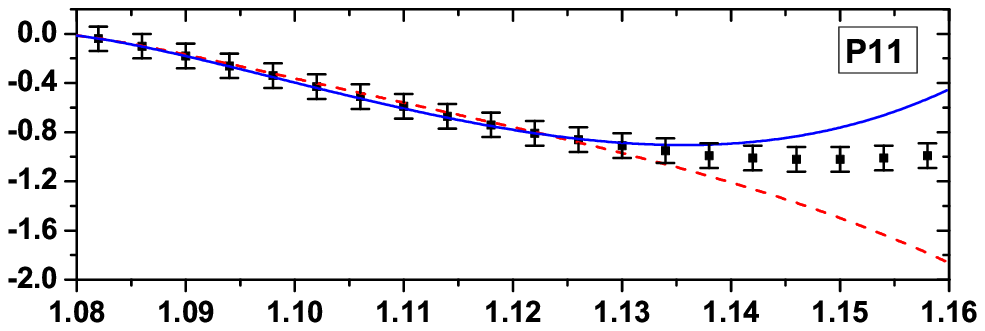}
%\vspace{-1.5cm}
\includegraphics[width=0.5\textwidth]{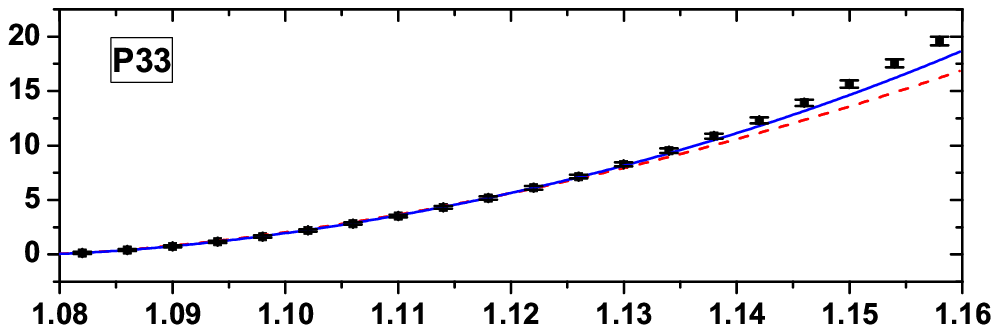}
%\vspace{-1.5cm}
\includegraphics[width=0.5\textwidth]{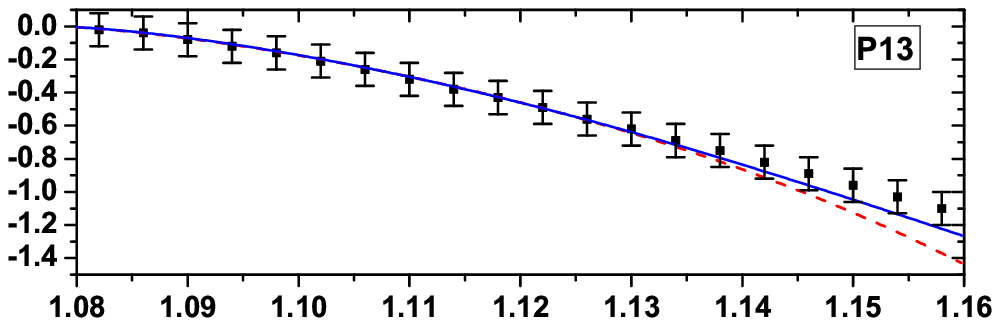}
\caption[pilf]{(Color online) Fit up to 1.13 GeV. The \textcolor{blue}{fourth-} and
\textcolor{red}{third-order} fits are presented by the
solid(\textcolor{blue}{blue}) and dash(\textcolor{red}{red}) lines
respectively.}\label{fig1}
\end{figure*}

\begin{figure*}[ht]
\includegraphics[width=0.5\textwidth]{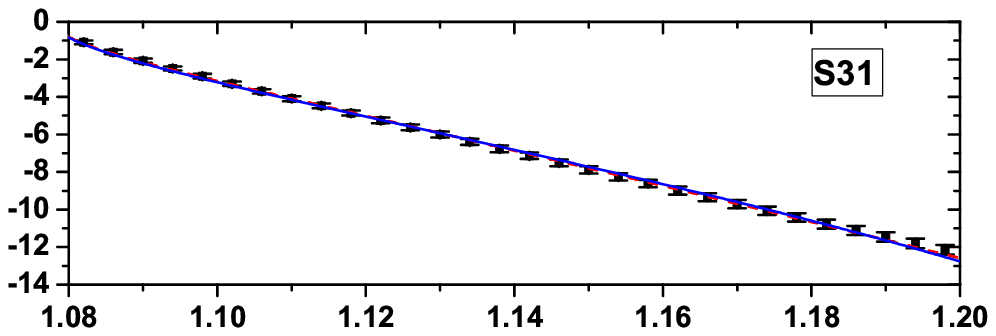}
\vspace{-1.0cm}
\includegraphics[width=0.5\textwidth]{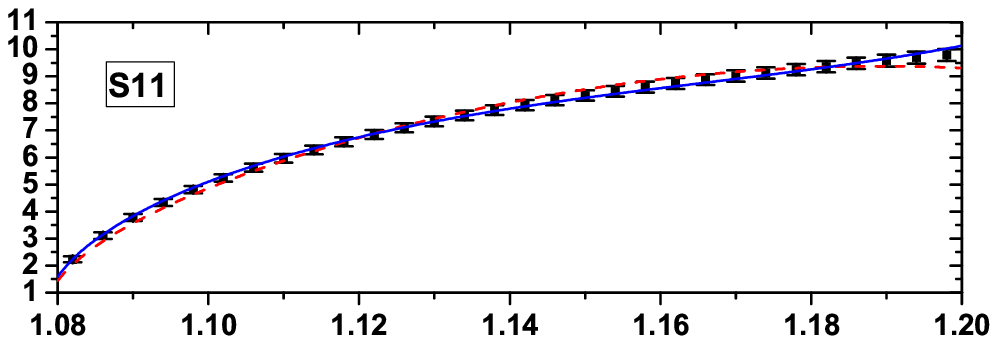}
\vspace{-1.0cm}
\includegraphics[width=0.5\textwidth]{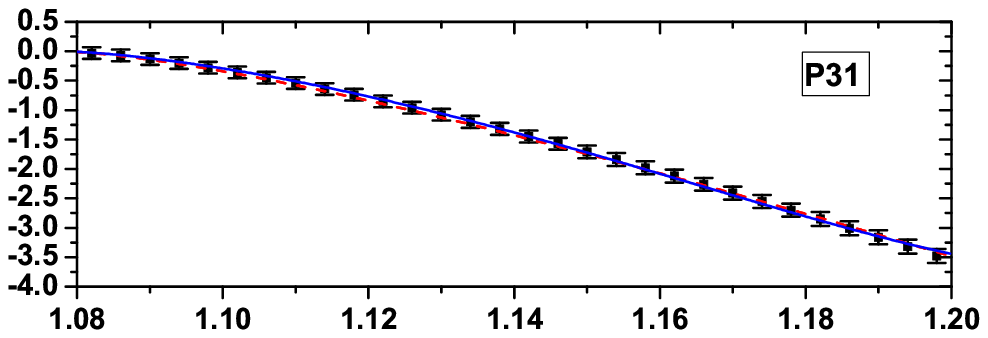}
\includegraphics[width=0.5\textwidth]{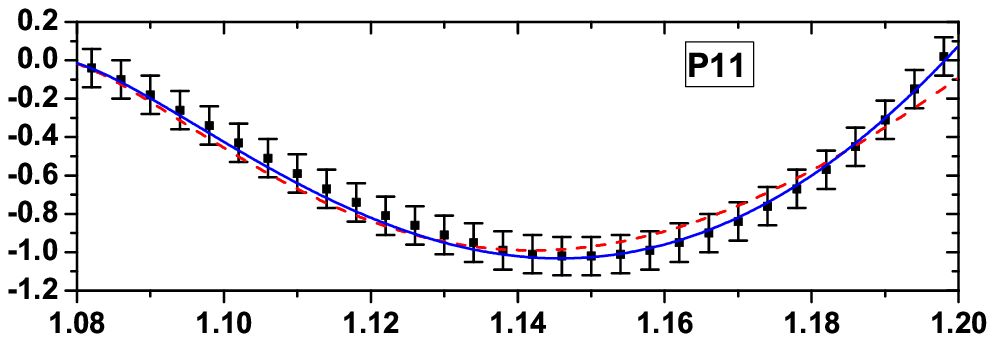}
\includegraphics[width=0.5\textwidth]{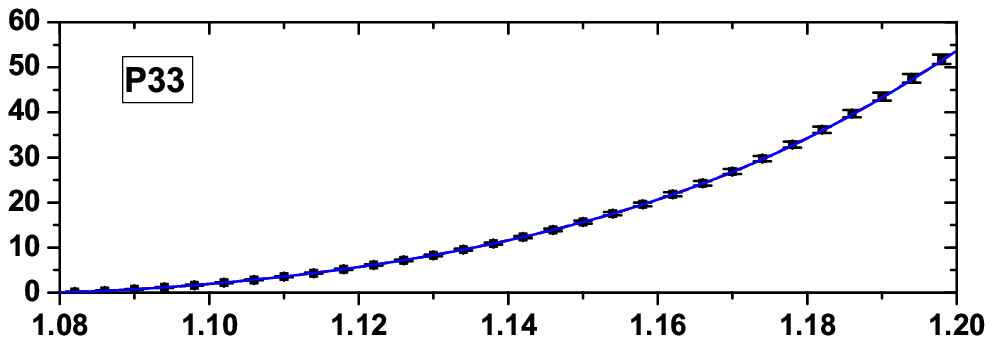}
\includegraphics[width=0.5\textwidth]{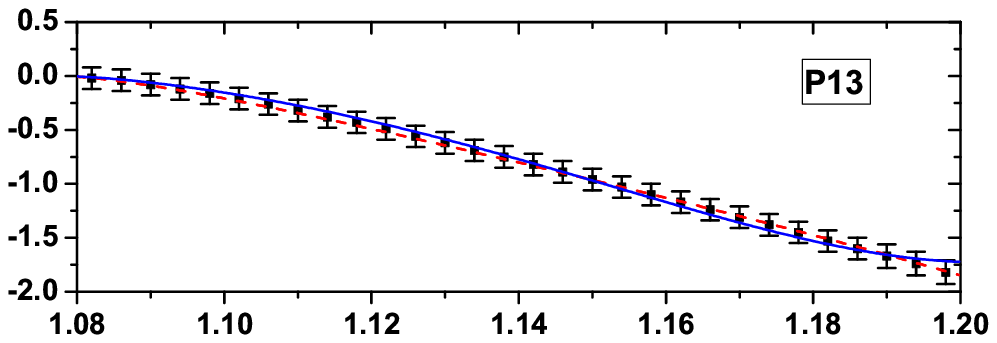}
\caption{(Color online) Fit up to 1.20 GeV. The \textcolor{blue}{fourth-} and
\textcolor{red}{third-order} fits are presented by the
solid(\textcolor{blue}{blue}) and dash(\textcolor{red}{red}) lines
respectively.}\label{fig2}
\end{figure*}

%%%%%%%%%%%%%%%%%%%%%%%%%%%%%%%%%%%%%%%%%%%%%%%%%%%%%%
\noindent
At $O(p^3)$ level we have performed two fits, the first one is up to
$\sqrt{s}=1.13$GeV, the second is up to $\sqrt{s}=1.20$GeV for the
convenience of comparing with the numerical studies given in
Ref.~\cite{oller1, oller2, oller3}. Data being fitted are from
Ref.~\cite{GW08} and error are assigned with the method of Ref.~\cite{oller2} . For the second fit we also included the tree level
$\triangle(1232)$ contribution~\cite{pascalutsa}, characterized by the $N\Delta$ axial coupling $h_A$. Fit results are
summarized in Table~\ref{table1}, where we have also listed the
results from Refs.~\cite{oller1} and \cite{oller3} for comparison.
We see that,  in general, our fit results at $O(p^3)$ level are in
good agreement with that of  Refs.~\cite{oller1,oller3}, except the
$d_5$ parameter. We also listed our $O(p^4)$ results from the best
solution in our fits. To let the fitted LECs same as~\cite{Fettes2000xg}, $d_{18}$ and $h_A$ are fixed at their $O(p^3)$ fitting results.% LEC are of natural size in the solution shown above.
 In Figures~\ref{fig1} and \ref{fig2} we plot the fit up to
$\sqrt{s}=1.13$GeV and 1.20GeV, respectively. We find that,
both $O(p^3)$ and $O(p^4)$ calculations give a reasonable
description to data and the $O(p^4)$ calculation improves the fit
quality.
%%%%%%%%%%%%%%%%%%%%%%%%%%%
%--------------------------Numerical fits-------------------------------------------------
%\foilhead[-0.8in]{\color{mycrimson}Numerical fits}

%%%%%%%%%%%%%%%%%%%%%%%%%%%

%%%%%%%%%%%%%%%%%%%%%%%%%%%
\section*{Acknowledgements}
\nin
%%%%%%%%%%%%%%%%
 We would like to thank Li-sheng Geng for helpful discussions. This work is supported in
part by National Nature Science
Foundations of China under contract number  10925522 %jieqing
%10875001%mianshang
 and
11021092.%tuandui
\appendix
\section{Definition of loop integrals}
\label{DefInt}
\nin
{\small
\begin{itemize}
\item
1 meson:
$\Delta_\pi=I_{10}$
\bea
\hspace{-20pt}\Delta_\pi=\intd\frac{1}{M^2-k^2}. \nonumber
\eea

\item
1 nucleon:
$\Delta_N=I_{01}$
\bea
\hspace{-20pt}\Delta_N=\intd\frac{1}{m^2-k^2}. \nonumber
\eea

\item
1 meson,1 nucleon:$I=I_{11}$
\bea
\hspace{-20pt}&&\{I,I^\mu,I^{\mu\nu}\}=\intd\frac{\{1,k^\nu,k^\mu k^\nu\}}{[M^2-k^2]\;[m^2-(\Sigma-k)^2]},\nonumber\\
\hspace{-20pt}&&I^\mu(s)=\Sigma^\mu I^{(1)}(s),\nonumber\\
\hspace{-20pt}&&I^{\mu\nu}(s)=g^{\mu\nu}I^{(2)}(s)+\Sigma^\mu\Sigma^\nu
I^{(3)}(s).\nonumber
\eea
%where
%\bea
%\hspace{-20pt}I^{(1)}(s)&=&\frac{1}{2s}\left\{\Delta_\pi-\Delta_N+(s-m^2+M^2)I(s)\right\},\nonumber\\
%\hspace{-20pt}I^{(2)}(s)&=&\frac{1}{d-1}\left\{M^2I(s)-\frac{1}{2}\Delta_N-\frac{1}{2}(s-m^2+M^2)I^{(1)}(s)\right\}.\nonumber
%\eea

\item
2 nucleons:$J_N=I_{02}$
\bea
\hspace{-20pt}J_N=\intd\frac{1}{[m^2-(k-P)^2]\;[m^2-(k-P^\prime)^2]}.\nonumber
\eea

\item
1 mesons,2 nucleon:$I_A$
\bea
\hspace{-20pt}I_{A}=\intd\frac{1}{[M^2-k^2]\;[m^2-(P-k)^2]\;[m-(P^\prime-k)^2]}.\nonumber
\eea
\end{itemize}}
After removing part proportional to $R=-\frac{1}{\epsilon}+\gamma_E-1-\ln4\pi$, the remaining scalar integrals are finite and denoted by, e.g. $\bar{I}(s)$, $\bar{J}_N(t)$, $\bar{I}_A(t)$, etc..

%% References


%\bibliographystyle{elsarticle-num}





% \end{thebibliography}

%%%%%%%%%%%%%%%%%%%%
%\vfill\eject

%%%%%%%%%%%%%%%

%\vfill\eject
\end{document}